\documentstyle[dcolumn,twoside,fleqn,espcrc2,epsf,graphicx]{article}

\newcolumntype{d}[1]{D{.}{.}{#1}}
\newcolumntype{e}[1]{D{,}{}{#1}}

\newcommand{\strace}{\mbox{$\:\mathsf{sTr}$}}
\newcommand{\ms}{m_{S}}
\newcommand{\mv}{m_{V}}
\def\({\left(}
\def\){\right)}
\def\[{\left[}
\def\]{\right]}

\title{Is the Up Quark Massless?}
\author{Daniel Nelson, George Fleming, and Greg Kilcup\address{Department of
Physics, The Ohio State University, Columbus, OH, 43210, USA}}

\begin{document}

\begin{abstract}
Several lattice calculations of a combination of the low energy constants of
the chiral Lagrangian, $2\alpha_8-\alpha_5$, are presented.  This combination
is critical for the preclusion of a massless up quark.  The result found is
$2\alpha_8-\alpha_5 = 0.115 \pm 0.051^{\mathrm stat} \pm 0.25^{\mathrm syst}$,
which is well outside of the range allowed by a massless up quark.
\end{abstract}

\maketitle

\section{Introduction}
The symmetries of QCD allow for a CP breaking term in the QCD Lagrangian.
However, measurements of the neutron dipole moment have demonstrated that the
coefficient of this term is very small, if not zero.  The unnatural smallness
of this coefficient is known as the strong CP problem.  A massless up quark has
long been proposed as a potential elegant solution to the problem.  Without the
up quark mass term, one is free to remove the CP violating term through a field
redefinition.  

At tree level Chiral Perturbation Theory (ChPT) appears to answer the question
of a massless up quark.  The quark mass ratios dictate the form of the
Lagrangian's mass matrix, which \emph{at lowest order} fully determines the
light meson mass ratios.  

However, the NLO terms in the chiral Lagrangian contribute to these ratios as
well.  If this contribution is strong enough, it would mimic the effects of a
massive up quark and allow a massless up quark to be consistent with
experimental results.  This is known as the Kaplan-Monohar ambiguity
\cite{KMambi}.

Distinguishing between a light and a massless up quark requires knowledge of
the coefficients of the NLO terms in the chiral Lagrangian, the
Gasser-Leutwyler (GL) coefficients.  Specifically, it is the combination of
constants $2\alpha_8 - \alpha_5$\footnote{The low energy constants $\alpha_i$
are related to the corresponding constants of the chiral Lagrangian $L_i$ by
the relation $\alpha_i = 8(4\pi)^2L_i$.} which corrects the relevant ratio.  If
this combination falls within a certain range, $-3.3 < 2\alpha_8 - \alpha_5 <
-1.5$, current experimental results can not rule out a zero up quark mass.

\begin{figure}
\begin{center}
\includegraphics[angle=0, width=0.465\textwidth]{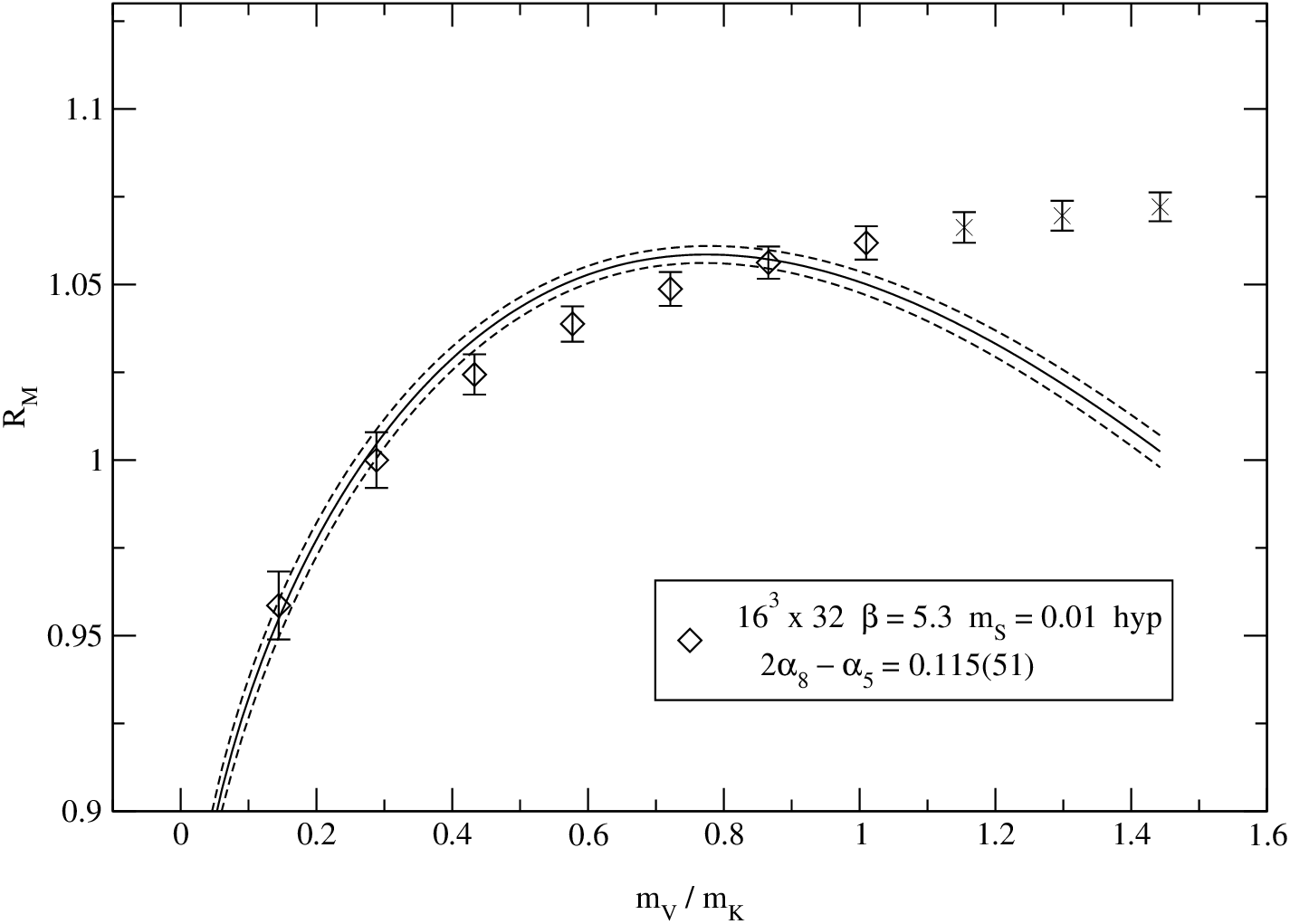}
\vspace{-14mm}
\caption{\small $16^3 \times 32$, $\beta=5.3$, $\ms = 0.01$, hyp;
$2\alpha_8-\alpha_5 = 0.115(51)$.  The $\times$'s were excluded from the fit.}
\label{figure:hyp_16_ratio}
\end{center}
\vspace{-10mm}
\end{figure}

The full set of GL coefficients, including the combination of interest, can not
be determined by experiment.  Various phenomenological arguments have been used
in the past to assemble a standard set of values for the coefficients.
However, these coefficients are determined by the low energy non-perturbative
behavior of QCD, and thus the lattice offers the best opportunity for
calculating them directly.

\begin{table}
\small
\setlength{\tabcolsep}{0.1pc}
\newlength{\digitwidth} \settowidth{\digitwidth}{\rm 0}
\catcode`?=\active \def?{\kern\digitwidth}
\caption{\small Simulation details.}
\begin{tabular}{|c|c|d{1.4}|d{1.3}|c|e{4.5}|c|d{7}|}
\hline
$L$ & $T$ & \multicolumn{1}{c|}{$\beta$} & \multicolumn{1}{c|}{$\ms$} & $N_f$ & \multicolumn{1}{c|}{$a^{-1}$ (MeV)\footnotemark} & hyp\footnotemark & \multicolumn{1}{c|}{$2\alpha_8-\alpha_5$} \\
\hline \hline
16 & 32 & 5.3   & 0.01  & 3 & 1347,.9(31) & $\surd$ & 0.115(51) \\
\hline
16 & 32 & 5.3   & 0.01  & 3 & 1260,(28)   &         & -0.071(14) \\
\hline
8 & 32 & 5.3    & 0.01  & 3 & 1366,(49)   &         & 0.47(12) \\
\hline
8 & 32 & 5.115  & 0.015 & 3 & 730,(130)   &         & 0.310(19) \\
\hline
8 & 32 & 5.1235 & 0.02  & 3 & 703,(97)    &         & 0.314(17) \\
\hline
8 & 32 & 5.132  & 0.025 & 3 & 780,(120)   &         & 0.375(13) \\
\hline
8 & 32 & 5.151  & 0.035 & 3 & 800,(130)   &         & 0.437(18) \\
\hline
\end{tabular}
\label{table:simulation_details}
\vspace{-6mm}
\end{table}

\section{Partially Quenched Chiral Perturbation Theory (PQChPT)} PQChPT is the
tool with which one can calculate the GL coefficients on the lattice.  PQChPT
is distinct from standard ChPT in that it is constructed from the symmetry of a
graded group.  This graded group follows from the presumed quark content of
Partially Quenched QCD (PQQCD), separate valence and sea quark flavors in
addition to ghost quark flavors, which in perturbation theory cancel loop
corrections due to valence quarks.

The Lagrangian of PQChPT up to $O(p^4)$ follows, with only relevant NLO terms
shown.
\begin{eqnarray}
\mathcal{L} &=& \frac{f^2}{4} \strace \[ \partial_\mu U \partial^\mu U^\dagger \] + \frac{f^2}{4} \strace \[ \chi U^\dagger + U \chi \] \nonumber \\
&+& L_4 \strace \[ \partial_\mu U \partial^\mu U^\dagger \] \strace \[ \chi U^\dagger + U \chi \] \nonumber \\
&+& L_5 \strace \[ \partial_\mu U \partial^\mu U^\dagger \( \chi U^\dagger + U \chi \) \] \nonumber \\
&+& L_6 \strace \[ \chi U^\dagger + U \chi \]^2 \nonumber \\
&+& L_8 \strace \[ \chi U^\dagger \chi U^\dagger + U \chi U \chi \] + \: \cdots
\end{eqnarray}
\begin{eqnarray}
& & U = \exp\( 2 i \Phi / f \) \\
& & \chi = 2 \mu \: \mbox{diag} \( \left\{ \ms, \mv \right\} \),
\end{eqnarray}
where $\Phi$ contains the pseudo-Goldstone ``mesons'' of the broken $SU(N_f +
N_V | N_V)_L \otimes \linebreak[0] SU(N_f + N_V | N_V)_R$ symmetry and $U$ is
an element of that group.  In our calculations three degenerate sea quarks were
used, $N_f = 3$, while the number of valence quarks $N_V$ cancels in all
expressions, affecting only the counting of external states.  The constants
$f$, $\mu$, and $L_i$ are unknown, determined by the low energy dynamics of
PQQCD.

It was recently realized that, because the valence and sea quark mass
dependence of the PQChPT Lagrangian is explicit and because full QCD is within
the parameter space of PQQCD ($\mv = \ms$), the values obtained for the GL
coefficients in a PQQCD calculation are the exact values for the coefficients
in full QCD \cite{Sharpe}\cite{Kaplan}.  Furthermore, the independent variation
of valence and sea quark masses allows additional lever arms in the
determination the these coefficients.  Because the $N_f$ dependence of the
Lagrangian is not explicit, the GL coefficients are functions of $N_f$.  Thus,
it is important to use a physical number of sea quarks, as we have, when
extracting physical results.  

\footnotetext[2]{Lattice spacing determined via $r_0$.}
\footnotetext[3]{Denotes a hypercubic blocked ensemble.}

\section{Predicted Forms}
PQChPT predicts forms for the dependence of the pseudoscalar mass and decay
constant on the valence quark mass, here assuming degenerate sea quarks and
degenerate valence quarks, and cutting off loops at $\Lambda_\chi = 4 \pi f$.
\begin{eqnarray}
m_\pi^2 &=& (4 \pi f)^2 z \mv \left\{ 1 \linebreak[0] + z \mv \( 2 \alpha_8 - \alpha_5 + \frac{1}{N_f} \) \right. \nonumber \\
&+& \left. \frac{z}{N_f} \( 2 \mv - \ms \) \ln z \mv \right\} \\
f_\pi &=& f \left\{ 1 + \frac{\alpha_5}{2} z \mv \right. \nonumber \\
&+& \left. \frac{z N_f}{4} \( \mv + \ms \) \ln \frac{z}{2} \( \mv + \ms \) \right\} \\
& & z = \frac{2 \mu}{(4 \pi f)^2}
\end{eqnarray}
These forms differ slightly from those in \cite{Sharpe2}, as the NLO dependence
in the sea quark mass has been absorbed into $\mu$ and $f$.  This is allowed as
the error due to this change manifests when $z$ appears in the NLO terms,
pushing the discrepancy up to NNLO.  Accounting for these absorbed terms would
require a systematic study at several sea quark masses.

\begin{figure}
\begin{center}
\includegraphics[angle=0, width=0.465\textwidth]{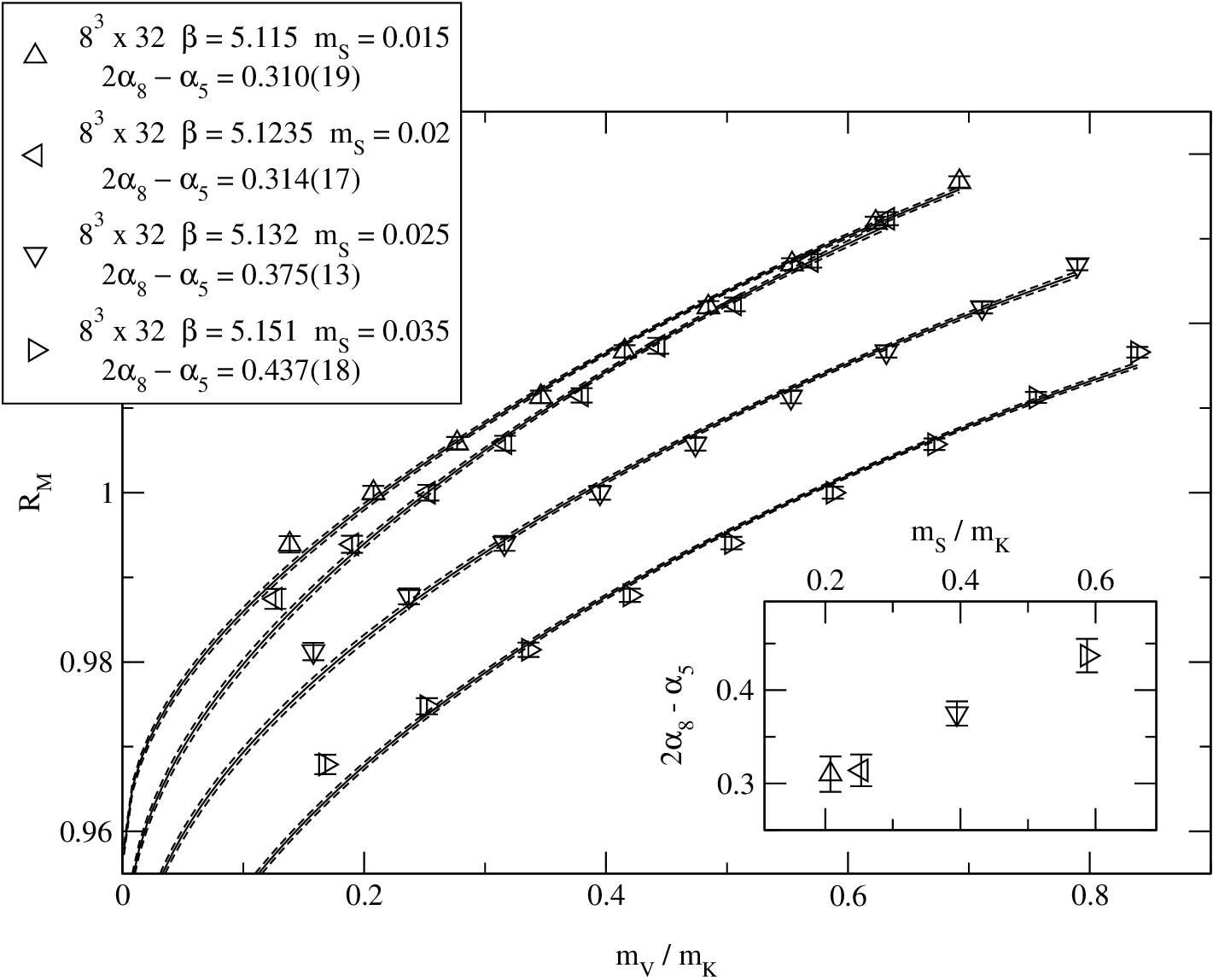}
\vspace{-14mm}
\caption{\small $8^3 \times 32$, several values of $\beta$ and $\ms$ with
matched lattice spacing.}
\label{figure:8x4_ratio}
\end{center}
\vspace{-10mm}
\end{figure}

The forms above are derived assuming degenerate light mesons.  However, our use
of staggered fermions on rather coarse lattices is likely to generate
significant flavor symmetry breaking, and thus a splitting of the light meson
masses.  Although there is no simple quantitative method to account for this
split, one effect might be a reduction in the importance of meson loops, and
thus a weakening of the chiral log terms.  In order to study the systematic
errors due to flavor symmetry breaking, we applied hypercubic blocking
\cite{hypblock} to our primary ensemble.

To determine the value of $2\alpha_8-\alpha_5$, the local pseudoscalar
correlator was calculated using several valence quark masses.  The correlators
were fit to exponentials, while simultaneously the meson mass and decay
constants were fit to the predicted forms.  For the smaller $8^3 \times 32$
ensembles, a constant term was added to the meson mass form, consistent with
finite volume.  The results of the fit are the values $\mu$, $f$, $\alpha_5$,
and $2\alpha_8-\alpha_5$.

Figure \ref{figure:hyp_16_ratio} and \ref{figure:8x4_ratio} display the quantity
\begin{eqnarray}
& & R_M = \frac{m^2_\pi(\ms)\mv}{m^2_\pi(\mv)\ms}
\end{eqnarray}
suggested by \cite{Alpha0}.  This quantity accentuates the NLO terms in
$m^2_\pi$, and thus is useful for plotting purposes.  It should be noted,
however, that the full forms of $m^2_\pi$ and $f_\pi$ were used when fitting.
When calculating the ratio, we did not use the simplification shown in
\cite{Alpha0}, but rather used a full numerator and denominator.  The value
$m_K$ in the figures is the valence quark mass at which the meson mass equals
the physical kaon mass.

\begin{figure}
\begin{center}
\includegraphics[angle=0, width=0.465\textwidth]{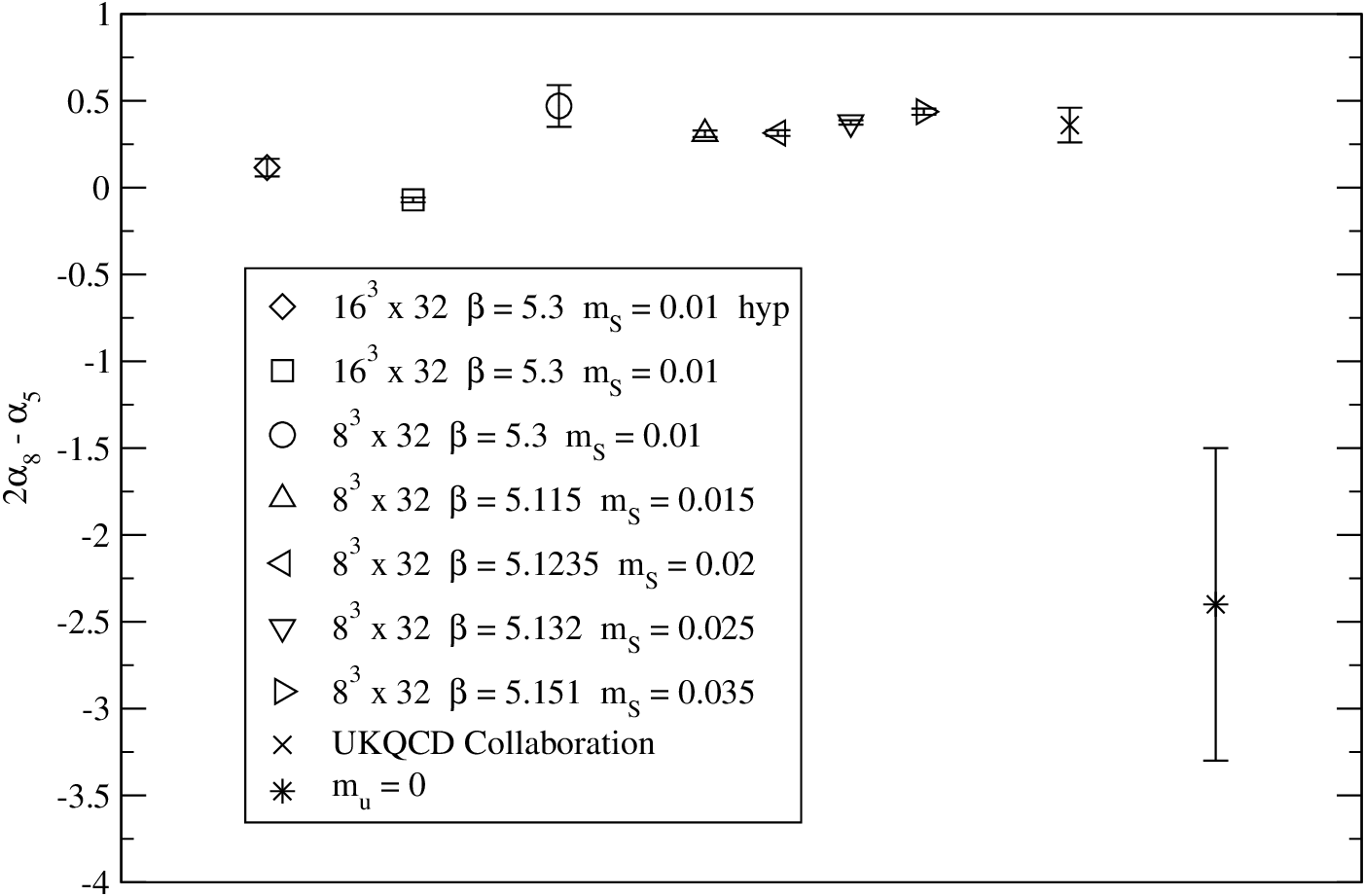}
\vspace{-14mm}
\caption{\small Compiled results.  The UKQCD Collaboration data point is taken
from \cite{Alpha}, showing their statistical error only.  The rightmost point
shows the range allowed by a massless up quark.}
\label{figure:results}
\end{center}
\vspace{-10mm}
\end{figure}

\section{Results}
$2\alpha_8-\alpha_5$ was found to vary significantly among the ensembles.
Changes in volume altered the result, suggesting that yet larger volumes may be 
required.  Additionally, the application of hypercubic blocking shifted the
result, suggesting that flavor symmetry breaking is indeed affecting the value.
Despite systematic shifts in the calculated value between ensembles, it remained
well outside the allowed range for a zero up quark mass.  

The quoted result of $2\alpha_8-\alpha_5 = 0.115 \pm 0.051^{\mathrm stat} \pm
0.25^{\mathrm syst}$ comes from our hypercubic blocked $16^3 \times 32$
ensemble, with generous systematic errors due to the significant fluctuations
in the result between ensembles.  This falls outside the range allowed by a
zero up quark mass, $-3.3 < 2\alpha_8-\alpha_5 < -1.5$.

\end{document}